\documentclass[aps,prb,reprint,superscriptaddress]{revtex4-2}

\usepackage{gensymb}
\usepackage{placeins}

\usepackage{amsmath}
\usepackage{amsfonts}
\usepackage{amssymb}
\usepackage{relsize}
\usepackage{graphicx}
\usepackage{makecell}
\usepackage{float}
\usepackage{color}

\definecolor{dancyan}{rgb}{0, 0.55, 0.55}
\definecolor{danorange}{rgb}{1, 0.45, 0}

\begin{document}

\title{Effects of Temperature Fluctuations on Charge Noise in Quantum Dot Qubits}
\author{Dan Mickelsen}
\author{Herve Carruzzo}
\affiliation{University of California, Irvine, California 92697, USA}
\author{S.~N.~Coppersmith}
\affiliation{University of New South Wales, Sydney, NSW 2052, Australia}
\author{Clare C. Yu}
\affiliation{University of California, Irvine, California 92697, USA}

\date{\today}

\begin{abstract}
Silicon quantum dot qubits show great promise but suffer from charge noise with a $1/f^\alpha$ spectrum, where $f$ is frequency and $\alpha \lesssim 1$.  It has recently been proposed that $1/f^\alpha$ noise spectra can emerge from a few thermally activated two-level fluctuators in the presence of sub-bath temperature fluctuations associated with a two-dimensional electron gas (2DEG)~\cite{Ahn2021}.  We investigate this proposal by doing Monte Carlo simulations of a single Ising spin in a bath with a fluctuating temperature.  We find that to obtain noise with a $1/f^\alpha$ spectrum with $\alpha \lesssim 1$ down to low frequencies, the duration of temperature fluctuations must be comparable to the inverse of the lowest frequency at which the noise is measured. This result is consistent with an analytic calculation in which the fluctuator is a two-state system with dynamics governed by time-dependent switching rates. In this case we find that the noise spectrum follows a Lorentzian at frequencies lower than the inverse of the average duration of the lowest switching rate. We then estimate relaxation times of thermal fluctuations by considering thermal diffusion in an electron gas in a confined geometry. We conclude that temperature fluctuations in a 2DEG sub-bath would require an unphysically long duration to be consistent with experimental measurements of 1/f-like charge noise in quantum dots at frequencies extending well below 1 Hz.
\end{abstract}

\newcommand{\beq}{\begin{eqnarray}}
\newcommand{\eeq}{\end{eqnarray}}
\maketitle

\section{Introduction}
Decoherence arising from charge noise presents a challenge to the use of silicon quantum dots (QDs) as quantum bits.  The charge noise spectrum in Si/SiGe quantum dots goes as $1/f^\alpha$ with $\alpha \lesssim 1$ over many decades in frequency~\cite{Connors2021,Jock2021}, where $f$ is the frequency and $\alpha$ is the noise exponent.  

Charge noise in quantum dots arises due to coupling to two-level fluctuators.  Experiments have shown that each quantum dot is coupled to a small number of fluctuators~\cite{Petit2020,Ahn2021}. While initial experiments indicated that the noise in neighboring quantum dots is not correlated ~\cite{Connors2019}, more direct subsequent experiments did find correlations~\cite{Yoneda2022}.  With only a few two-level fluctuators, a Lorentzian power spectra is expected, but instead, noise with a $1/f^\alpha$ power spectrum is observed with $\alpha \lesssim 1$. Typically, 1/f-like noise is produced by an ensemble of two-level fluctuators with a broad distribution of relaxation rates~\cite{Dutta1979,Dutta1981}.

Ahn {\it et al.} \cite{Ahn2021} have suggested that the quantum dots are coupled to a small number of two-level fluctuators that are each in turn coupled to a microscopic subsection of the larger thermal bath that they take to be the 2D electron gas (2DEG) in which the quantum dots are embedded.  They propose that temperature fluctuations in the sub-bath cause the noise to have a $1/f^\alpha$ spectrum over several decades of frequency with $\alpha\sim 1$. 
They calculate the noise power spectral density by performing a quenched average over a distribution of temperatures and show that this average yields a $1/f^\alpha$ noise spectrum even for small numbers of fluctuators.
However, they did not specify the conditions under which the quenched average is justified. (By ``quenched average'', we mean an average over a distribution of temperatures where each temperature is infinitely long lived.)

In this paper we show that the noise spectrum arising from a two-level fluctuator (TLF) in the presence of a time-varying temperature is described by a quenched average only if the time scale of the temperature variations is longer than the lowest frequency measured.  We also estimate the time scale of temperature variations in typical silicon qubit devices and find that it is likely to be substantially faster than the lowest frequencies at which a $1/f^\alpha$ noise spectrum has been measured in qubit devices~\cite{Connors2019}.
%

To illustrate these points, we represent the fluctuator by a single Ising spin with thermally activated flips subjected to temperature fluctuations. Using Monte Carlo simulations of this spin, we find that a 1/f magnetic noise spectrum requires very slow temperature fluctuations, with each temperature being extremely long-lived. To confirm this result, we perform an analytical calculation where the fluctuator is a two-state system that has a time dependent switching rate. We consider the case where the switching rate is a sequence of constant, but random, rates. Each rate has an average duration $t_o$ and corresponds to a certain temperature. Changing the rate corresponds to changing the temperatures. We find that $t_o$ must be very long in order to achieve 1/f noise at low frequencies. We use our result to reproduce the results of Ahn {\it et al.} where $t_o$ is infinite and show what happens when the temperature fluctuations have a finite lifetime. We then estimate the longest possible time scale of temperature fluctuations based on thermal diffusion in the confined geometry of the device and conclude that the temperature fluctuations in a sub-bath of the 2DEG cannot live long enough to account for the observed 1/f noise.

\section{Monte Carlo Simulation of a Single Ising Spin}
In the model of Ahn {\it et al.} a thermally activated two-level fluctuator ~\cite{Ahn2021} is coupled to a quantum dot, with fluctuations in the TLF leading to charge noise in the quantum dot. The thermally activated time for the fluctuator to switch is given by $\tau\exp(U/T)$ where $U$ is barrier height and $\tau^{-1}$ is the attempt frequency.
While a single TLF with a characteristic temperature-dependent relaxation rate produces a Lorentzian noise power spectrum, if the TLF is coupled to a microscopic subsection of the larger thermal bath, as Ahn \emph{et al}.~propose~\cite{Ahn2021}, then fluctuations in the sub-bath temperature have the potential to give rise to multiple relaxation rates associated with a single TLF, resulting in a 1/f noise power spectrum.
Here, we wish to elucidate how the noise spectrum depends on the nature of the time variation of the temperature fluctuations of the bath.

To incorporate the time-dependence of the bath, our simulations represent the thermally activated TLF by a single Ising spin which flips between $S_z=-1$ and $S_z=+1$ with a probability $P_i=e^{-U/T_i}$ where $U=1$ and $T_i$ is the temperature during the $i$th time interval, and time is measured in units of $\tau$. Since the temperature depends on time, there is a sequence of different temperatures that are drawn from a Gaussian distribution that is truncated to exclude negative temperatures~\cite{Ahn2021}:
\begin{equation}
\label{eq:AhnTdist1}
f(T_{i},T_{\rm avg},\sigma_\text{sb})=\xi\frac{e^{-(T_i-T_{\rm avg})^2/2\sigma_\text{sb}^2}}{\sqrt{\pi/2}\sigma_\text{sb}},
\end{equation}
where $T_{\rm avg}$ is the average sub-bath temperature, $\sigma_\text{sb}^2$ the variance, and $\xi$ is a normalization factor that accounts for the truncation of the Gaussian. This distribution is the one used by Ahn {\it et al.} \cite{Ahn2021}. Following Ahn {\it et al.}, we set $T_{\rm avg}=1$ and $\sigma_\text{sb}=0.3$. In this case $\xi\sim 1$.

The spin is reoriented according to standard Monte Carlo dynamics. A random number between zero and 1 is generated from a uniform distribution and if it is less than or equal to $P_i$, the spin flips. The duration $\Delta_i$ of the $i$th temperature is drawn from an exponential distribution:
\begin{equation}
P_d(\Delta)=\frac{1}{t_o}e^{-\Delta/t_o},
\label{eq:DeltaDist}
\end{equation}
where $t_o$ is the characteristic duration of a given temperature. The length of each run is $6\times 10^8$ time steps. Noise power spectra are calculated during the runs at 30 frequencies evenly spaced on a logarithmic scale.

\begin{figure}
\centering
\includegraphics[width=\linewidth]{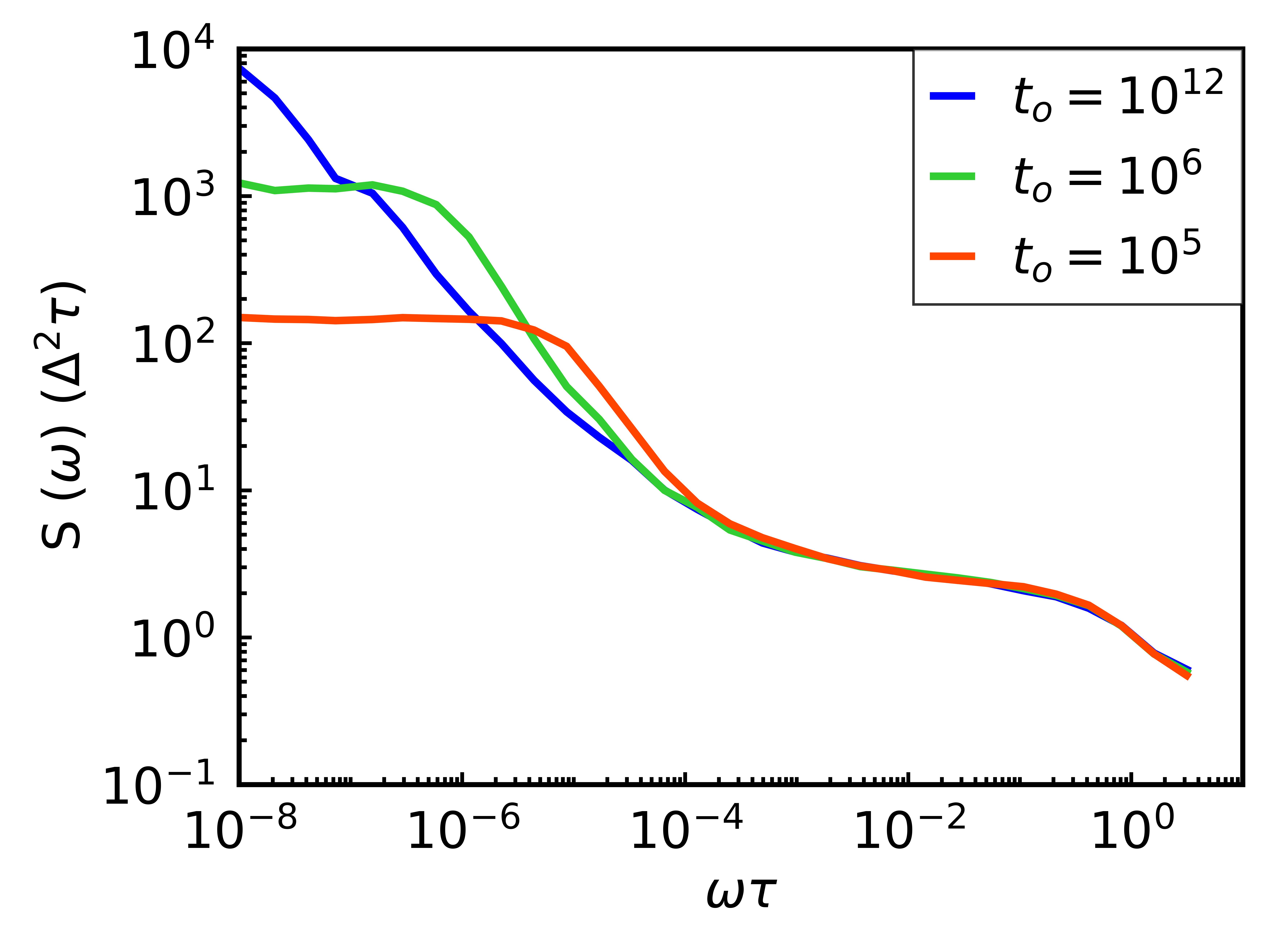}
\caption{Plot of the spectral power densities of the noise, $S(\omega)$, versus $\omega \tau$, where $\omega$ is the angular frequency and $1/\tau$ is the attempt frequency, for an Ising spin in the presence of temperature fluctuations with average duration $t_o=10^5$ (red), $10^6$ (green), and $10^{12}$ (blue) (as defined in Eq.~(\ref{eq:DeltaDist})). All values of $t_0$ are measured in units of $\tau$.
The noise spectra are averaged over 13 sets of 300 runs. Each run consists of $6\times 10^8$ Monte Carlo Steps (MCS). 
These results demonstrate that the noise spectral power density saturates at low frequencies, with the saturation frequency decreasing as $t_0$ increases. 
}
\label{IsingSpinSpectra}
\end{figure}

The low frequency noise spectrum is dominated by low temperatures that are in the tail of the distribution and are, therefore, not often sampled by random draws. Therefore, we divide the temperature range from $T=0.03$ to $T=2.0$ into 300 equal increments and start one run from each of these 300 temperatures. After the initial temperature of a run is finished, i.e., after $\Delta_1$ steps, the subsequent temperatures in that run are chosen randomly from the Gaussian distribution. The noise spectrum from the resulting time series is given a Gaussian weight corresponding to the initial temperature and Eq. (\ref{eq:AhnTdist1}). This is how we average over the 300 noise spectra. 13 sets of 300 runs were performed. The results are shown in Fig.~\ref{IsingSpinSpectra} for $t_o=10^5,\;10^6,\;10^{12}$. All the spectra are normalized so that the total noise power is unity. One can see that as $t_o$ increases, the knee moves to lower frequencies indicating that slow, long-lived temperature fluctuations are needed to observe 1/f noise at low frequencies.

We can compare these power spectra to the one of Ahn \emph{et al}.~\cite{Ahn2021}. Ahn {\it et al.} had infinitely long-lived temperature fluctuations and so their results should be compared to our case of $t_o=10^{12}$. For shorter $t_o$, one can see that there is a low-frequency knee where the power spectra crosses over from white noise at low frequency to 1/f noise at higher frequencies.  The frequency at which this knee occurs is proportional to $1/t_o$.
Our contention that the temperature fluctuations must be slow in order for the noise to have a 1/f spectrum at low frequencies is supported by our power spectra, where increasing $t_o$ lowers the frequency of the knee. This knee is not present in the power spectrum of Ahn \emph{et al}.
because their calculation was a quenched average over an ensemble of TLFs, each remaining at a different temperature forever. 

\section{Sequence of Random Fluctuation Rates}
To confirm our finding that a fluctuating temperature results in a 1/f-like noise spectrum only when the time scale of the thermal fluctuations are longer than the inverse of the lowest frequency at which 1/f noise is observed, we did an analytic calculation where we consider a two state fluctuator with a time-dependent transition rate $\gamma^{\prime}(t)$ between two degenerate minima at $x=0$ and $x=1$. The correlation function $\Psi(t)= \langle x(0)x(t) \rangle$ (where the angular brackets denote an average over  realizations selected from the specified distribution) is just half of $p_1(t)$, the probability of the system being in state 1 at time $t$, given that it was in state 1 at $t=0$. (The factor of 1/2 is the probability that the system is in state 1 at $t=0$.) The equation governing the evolution of $p_1(t)$ is:
\begin{equation}\label{eq:Revolution_p1}
\frac{dp_1(t)}{dt}=-\gamma^{\prime}(t) p_1(t) + \gamma^{\prime}(t)  p_0(t) = \gamma^{\prime}(t)(1- 2p_1(t)),
\end{equation}
where $p_0(t)=1-p_1(t)$ is the probability that the system is in state 0 at time $t$ and the initial condition is $p_1(0)=1$. The solution of this equation is:
\begin{equation}\label{eq:Revolution_sol}
p_1(t)=\frac{1}{2} \left ( 1+e^{-2\int_0^t\gamma^{\prime}(t_1)dt_1} \right )~.
\end{equation}
Since $\gamma^{\prime}(t) > 0$, the probability approaches 1/2 as $t\rightarrow\infty$, as it should in equilibrium when the states are degenerate. In the following, we rescale $\gamma^{\prime}(t)$ such that $\gamma(t)=2\gamma^{\prime}(t)$. Since we are interested in fluctuations about the mean, we define the autocorrelation function accordingly:
\begin{eqnarray}
\label{eq:Revolution_sol2}
\Psi(t)&=&\langle (x(0)-\langle x \rangle) (x(t)- \langle x\rangle) \rangle\nonumber \\
&=& \langle x(0)x(t) \rangle - 1/4\nonumber\\
&=& \frac{1}{4}e^{-\int_0^t\gamma(t_1)dt_1},
\end{eqnarray}
where we used $\langle x \rangle =1/2$. 

The Fourier transform $\Psi(\omega)$ of the correlation function $p_1(t)/2$ is given by
\begin{equation}\label{eq:RFTrelfunc}
\Psi(\omega)=\frac{1}{2}\int_0^{\infty}e^{-\int_o^t\gamma(t')dt'}\cos(\omega t)dt,
\end{equation}
where we have used the symmetry $\Psi(t)=\Psi(-t)$. We can replace $\gamma(t)$ by a discrete sequence of constant rates such that the $n$th transition rate $\gamma_n$ occurs during the time interval $\Delta_n=t_{n}-t_{n-1}$. Writing $t_n=\sum_{i=1}^n\Delta_i$, the integral Eq. (\ref{eq:RFTrelfunc}) is broken into these time intervals:
\begin{equation}\label{eq:RFTrelfuncchunked}
  \begin{split}
    \Psi(\omega)&=\frac{1}{2}\int_0^{\infty}e^{-\int_o^t\gamma(t')dt'}\cos(\omega t)dt\\
    &=\frac{1}{2}\left(\int_0^{t_1}e^{-\gamma_1t}\cos(\omega t)dt\right.\\
    &\left.+\int_{t_1}^{t_2}e^{-\gamma_1\Delta_1-\gamma_2(t-t_1)}\cos(\omega t)dt+\ldots\right.\\
    &\left.+\int_{t_k}^{t_{k+1}}e^{-(\sum_{i=1}^k\gamma_i\Delta_i)-\gamma_{k+1}(t-t_k)}\cos(\omega t)dt+\ldots\right)\\
                &=\frac{1}{2}\sum_{k=0}^{\infty}e^{-(\sum_{i=1}^k\gamma_i\Delta_i)}\int_{t_k}^{t_{k+1}}e^{-\gamma_{k+1}(t-t_k)}\cos(\omega t)dt \\
    &=\frac{1}{2}\sum_{k=0}^{\infty}e^{-(\sum_{i=1}^k\gamma_i\Delta_i)}\\
    &\times\int_{0}^{\Delta_{k+1}}e^{-\gamma_{k+1}t}
    \cos(\omega t+\omega \sum_{i=1}^k\Delta_i)dt.
    \end{split}
\end{equation} 
This can be further simplified by expressing the cosine in terms of exponentials:
\begin{equation}\label{eq:RFTrelfuncchunked2}
  \Psi(\omega) = \textrm{Re}\left\{
    \sum_{k=0}^{\infty}e^{-(\sum_{i=1}^k(\gamma_i-i\omega)\Delta_i)}\int_{0}^{\Delta_{k+1}}e^{-(\gamma_{k+1}-i\omega)t}dt
  \right\}.
\end{equation}
The time integral can then be performed:
\begin{equation}\label{eq:RFTrelfuncchunked3}
  \Psi(\omega) = \frac{1}{2}\textrm{Re}\left\{
    \sum_{k=0}^{\infty}e^{-(\sum_{i=1}^k(\gamma_i-i\omega)\Delta_i)}\frac{1-e^{-(\gamma_{k+1}-i\omega)\Delta_{k+1}}}{\gamma_{k+1}-i\omega}
  \right\}.
\end{equation}
This expression for the frequency dependence of the noise must then be averaged over all possible realizations of $\Delta_i$ and $\gamma_i$ which we assume to be independent, i.e., $P(\Delta_1..\Delta_n\gamma_1..\gamma_n)=P_d(\Delta_1)\cdot ..\cdot P_d(\Delta_n)P_g(\gamma_1)\cdot ..\cdot P_g(\gamma_n)$, where $P_d(\Delta_i)$ and $P_g(\gamma_i)$ are the respective distributions of individual $\Delta_i$ and $\gamma_i$. We model $P_d(\Delta_i)$ as the arrival time in a queue, i.e. an exponential distribution:
\begin{equation}\label{eq:DeltaDistribution}
P_d(\Delta)=\frac{1}{t_o}e^{-\Delta/t_o},
\end{equation}
where $t_o$ is the mean duration of a given value of the relaxation rate, i.e., it is the mean time between changes in the relaxation rate. 
With this distribution, the average over the $\Delta_i$'s in Eq. (\ref{eq:RFTrelfuncchunked3}) can be done. The average over the distribution of $\gamma_i$ is left as a formal average $\langle .. \rangle_{\gamma}$ for now.  We can rewrite $\Psi(\omega)$ in Eq. (\ref{eq:RFTrelfuncchunked2}) as
\begin{equation}
      \Psi(\omega) = \frac{1}{2}\textrm{Re}\left\{
    \sum_{k=0}^{\infty}\mu_0^k(\omega)\mu_1(\omega)
  \right\},
 \label{eq:RFTcorrFactor}
\end{equation}
where $\mu_o$ is independent of $i$ since $P_d$ is the same for all $i$ as can be seen in the following expression:
\begin{eqnarray}
\mu_o(\omega)&=&\left\langle \frac{1}{t_o}\int_0^{\infty}e^{-(\gamma_i-i\omega)\Delta_i
-\Delta_i/t_o}d\Delta_i\right\rangle_{\gamma} \nonumber\\
&=&
\left\langle \frac{1}{1+\gamma t_o-i\omega t_o}\right\rangle_{\gamma}.
\label{eq:Rmu0}
\end{eqnarray}
Note also that $|\mu_0(\omega)|<1$. The average in $\mu_1(\omega)$ is given by:
\begin{equation}\label{eq:Rmu1}
  \begin{split}
    \mu_1(\omega)&=\left\langle \frac{1}{\gamma_{k+1}-i\omega}\cdot \right. \\&\quad\quad
    \left. \left( 1-\frac{1}{t_o}
    \int_0^{\infty}e^{-(\gamma_{k+1}-i\omega)\Delta_{k+1}-\Delta_{k+1}/t_o}d\Delta_{k+1}\right) \right\rangle_{\gamma}\\
                 &=\left\langle \frac{1}{\gamma -i\omega}\left(1-\frac{1}{1+\gamma t_o-i\omega t_o}\right)\right\rangle_{\gamma}\\
                 &=\left\langle \frac{t_o}{1+\gamma t_o-i\omega t_o}\right\rangle_{\gamma}\\
    &= t_o\mu_0(\omega)
  \end{split}
\end{equation}
which is independent of $k$. 
The geometric sum can now be carried out:
\begin{equation}\label{eq:RFTrelfuncchunked4}
  \begin{split}
  \Psi(\omega) &= \frac{t_o}{2}\textrm{Re}\left\{
    \sum_{k=0}^{\infty}\mu_0^k(\omega)\mu_0(\omega)
  \right\}\\
  &=
    \frac{t_o}{2}\textrm{Re}\left\{
    \frac{\mu_0(\omega)}{1-\mu_o(\omega)}
                 \right\}\\
    &=\frac{t_o}{2}\cdot \frac{\mu_0^{'}(\omega)-(\mu_0^{'}(\omega))^2-(\mu_0^{''}(\omega))^2}{1-2\mu_0^{'}(\omega)+(\mu_0^{'}(\omega))^2+(\mu_0^{''}(\omega))^2},
    \end{split}
\end{equation}
where we have used the definition $\mu_i(\omega)=\mu_i^{'}(\omega)+i\mu_i^{''}(\omega)$. Explicit formulas for $\mu_0$ from Eq. (\ref{eq:Rmu0}) for a discrete distribution of relaxation rates specified by $m$ values $\gamma_1..\gamma_m$, equally weighted for simplicity, are:
\begin{equation}\label{eq:Rmu0Re}
\mu_0^{'}(\omega)=\frac{1}{m}\sum_l\frac{1+\gamma_lt_o}{(1+\gamma_lt_o)^2+(\omega t_o)^2}
\end{equation}
\begin{equation}\label{eq:Rmu0Im}
\mu_0^{''}(\omega)=\omega t_o\frac{1}{m}\sum_l\frac{1}{(1+\gamma_lt_o)^2+(\omega t_o)^2},
\end{equation}
where the sum over $l$ runs from $1$ to $m$. Eq. (\ref{eq:RFTrelfuncchunked4}) together with Eqs. (\ref{eq:Rmu0Re}) and (\ref{eq:Rmu0Im}) is the main result of this section. It is a bit difficult to see the overall behavior in this expression since it depends on the values chosen for the $\gamma_l$ in the formulas for the averages entering the power spectra. However, the limit $t_o\rightarrow\infty$ is easy to obtain and gives
\begin{equation}\label{eq:spectraQuenchedLimit}
\Psi(\omega)=\frac{1}{2 m}\sum_l\frac{\gamma_l}{\omega^2+\gamma_l^2}
\end{equation}
which is the quenched limit. With appropriate values for $\gamma_l$, this easily produces a $1/\omega$ noise spectrum spanning decades in frequency.
\begin{figure}[h]
\centering
\includegraphics[width=8.5cm]{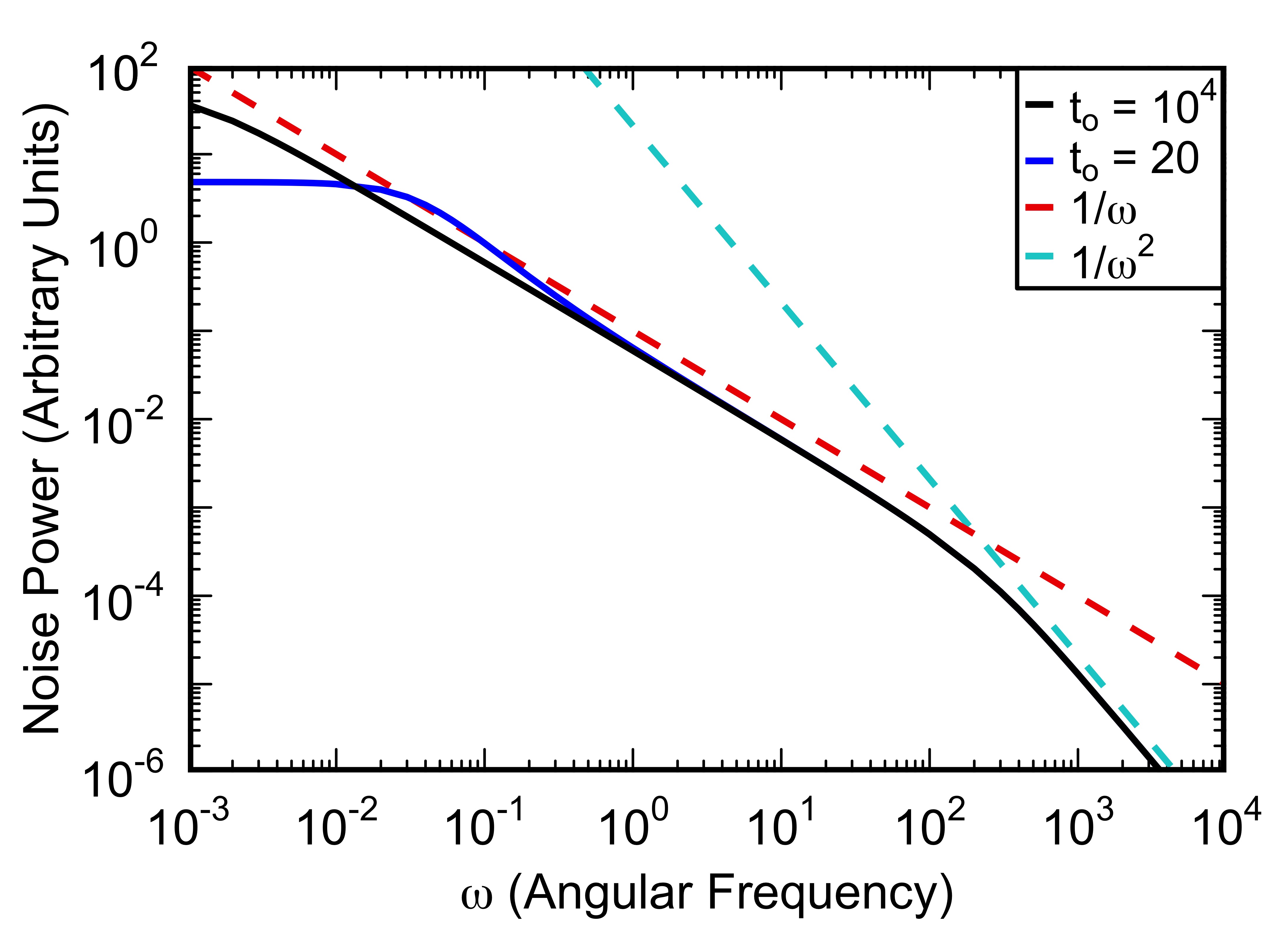}
\caption{Comparison of noise spectra on the time scale on which the transition rates vary, obtained using the analytic theory.   The plot shows the noise power vs.\ angular frequency on a log-log plot with a rate change on average every $t_o=20$ (blue solid line) and every $t_o=10^4$ (black solid line). Straight dashed lines are guides to the eye. The red dashed line corresponds to $1/\omega$ and the cyan dashed line corresponds to $1/\omega^2$. The discrete distribution of $\gamma$ is discussed in the text.  
The 1/f-like behavior of the spectrum is cut off at frequencies below $1/t_0$.
\label{fig:RandomFluctSpectraDiscrete}}
\end{figure}

More insight requires the specification of values for $\gamma_l$. Choosing  $\gamma_l$ to be $\gamma_l=2^l$ where $l$ = -10, -9,..., 7, 8, gives a fairly good approximation to a $1/\omega$ noise power spectrum over several decades in frequency when averaging over the Lorentzian noise spectra for each $\gamma_l$; this is the quenched limit. Fig.~\ref{fig:RandomFluctSpectraDiscrete} shows the resulting spectra for a large $t_o=10^4$ (black line) which is indistinguishable from the quenched average and can be seen to follow a $1/\omega$ power law (red line). At a much shorter switching time of $t_o=20$ (blue line), the power spectra turns flat at low frequencies sooner, when going from high to low frequencies, than in the quenched limit. One can see a short range of frequencies where the power spectra goes roughly as $1/\omega$ before going over to $1/\omega^2$ when $\omega$ exceeds the largest rate.

A continuous distribution for $\gamma$ can also be considered. For instance, $p_1(\gamma)=\lambda^{-1}/\gamma$ with $\lambda=\log(\gamma_\text{max}/\gamma_\text{min})$ and $\gamma_\text{min}<\gamma<\gamma_\text{max}$ will give a $1/\omega$ spectrum in the quenched limit. Using this distribution instead of the sums in Eqs. (\ref{eq:Rmu0Re}) and (\ref{eq:Rmu0Im}) yields:
\begin{eqnarray}\label{eq:muC}
\mu_0^{'}(\omega) &=&  D(\omega)-\omega t_o E(\omega) \nonumber\\
\mu_0^{''}&=&\omega t_o D(\omega)+E(\omega)
\end{eqnarray}
with
\begin{eqnarray}\label{eq:DandE}
D(\omega)&=&\frac{2\lambda -B(\omega)}{2\lambda C(\omega)} \nonumber\\
E(\omega) &=& \frac{A(\omega)}{2\lambda C(\omega)}
\end{eqnarray}
\begin{equation}\label{eq:A}
A(\omega)=2\left(\arctan\left(\frac{\gamma_\text{min} t_o+1}{\omega t_o}\right)-\arctan\left(\frac{\gamma_\text{max} t_o+1}{\omega t_o}\right)\right)
\end{equation}
\begin{equation}\label{eq:B}
B(\omega)=\log\left(\frac{\omega^2 t_o^2+\gamma_\text{max}^2t_o^2+2\gamma_\text{max} t_o+1}{\omega^2 t_o^2+\gamma_\text{min}^2t_o^2+2\gamma_\text{min} t_o+1}\right)
\end{equation}
and
\begin{equation}\label{eq:C}
C(\omega)= \omega^2 t_o^2+1~.
\end{equation}

\begin{figure}[h]
\centering
\includegraphics[width=8.5cm]{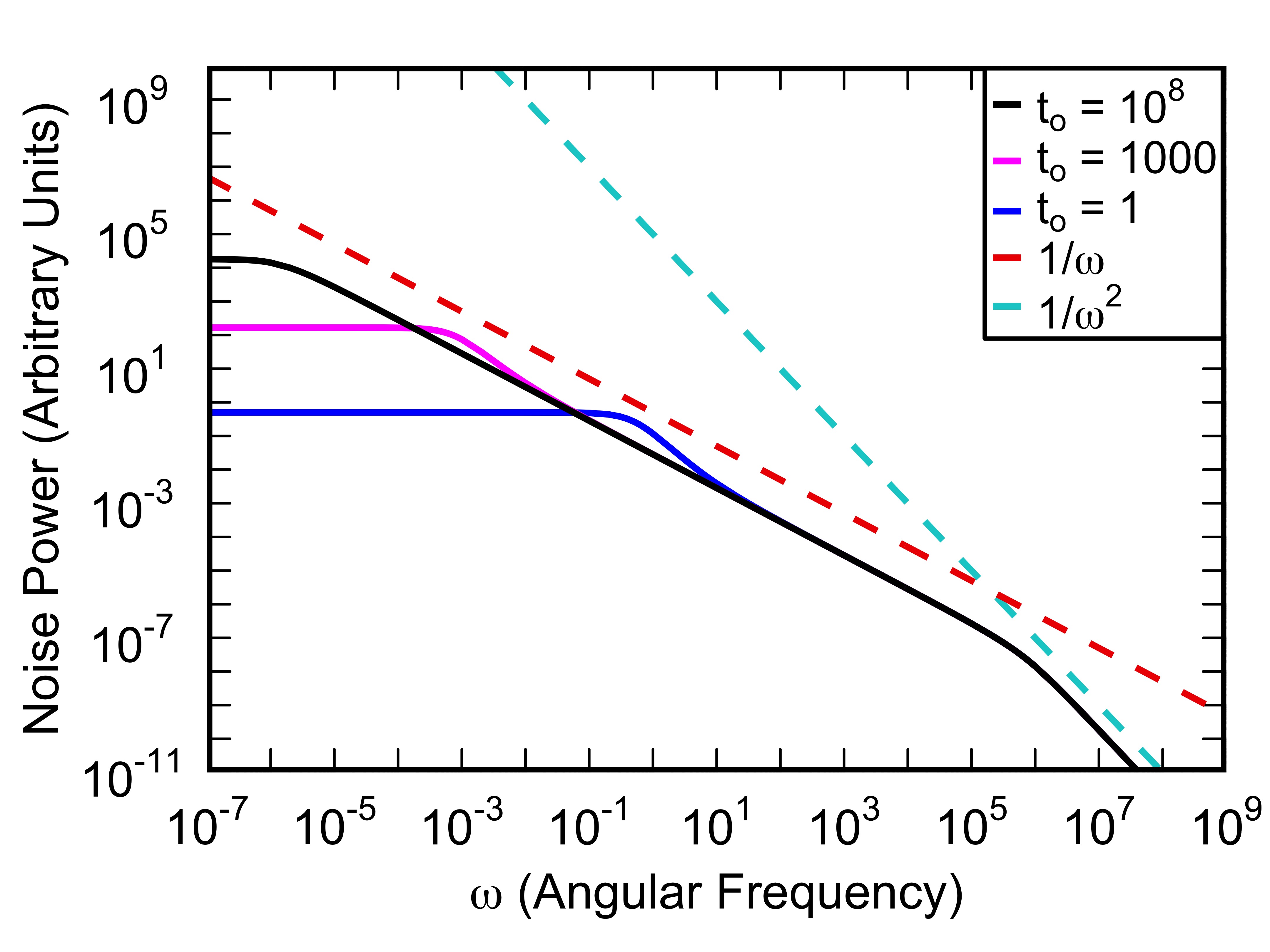}
\caption{Noise power spectra vs. angular frequency on a log-log plot for a continuous distribution of rates with a rate change on average every $t_o=10^8$ (black), $t_o=1000$ (magenta) and $t_o=1$ (blue) respectively. $\gamma_\text{min}$ is $10^{-6}$ and $\gamma_\text{max}$ is $10^6$. Straight dashed lines are guides to the eye. The red dashed line goes as $1/\omega$ and the cyan dashed line goes as $1/\omega^2$.
The spectrum is 1/f-like at intermediate frequencies when the characteristic time $t_0$ describing the rate variations is long, so that ${\rm max}[\gamma_\text{min},1/t_0]\ll\omega\ll\gamma_\text{max}$.}
\label{fig:RandomFluctSpectraCont}
\end{figure}

A plot of the power spectra is shown in Fig. \ref{fig:RandomFluctSpectraCont} for $\gamma_\text{min}=10^{-6}$ and $\gamma_\text{max}=10^{6}$. The black curve corresponds to a mean time of $t_o=10^8$ between rate changes. At frequencies below the lowest rate in the system, $10^{-6}$, the noise becomes flat. At frequencies higher than the largest rate, the response goes over to $1/\omega^2$. The magenta curve shows the effect of a faster mean switching time of $1000$, flattening at frequencies below $\omega\cdot 1000\sim 1$, instead of at the slowest rate of $10^{-6}$. At the still lower mean switching time of 1, shown by the blue curve, the crossover occurs at higher frequencies, again dictated by $\omega t_o\sim 1$.

\begin{figure}[h]
\centering
\includegraphics[width=8.5cm]{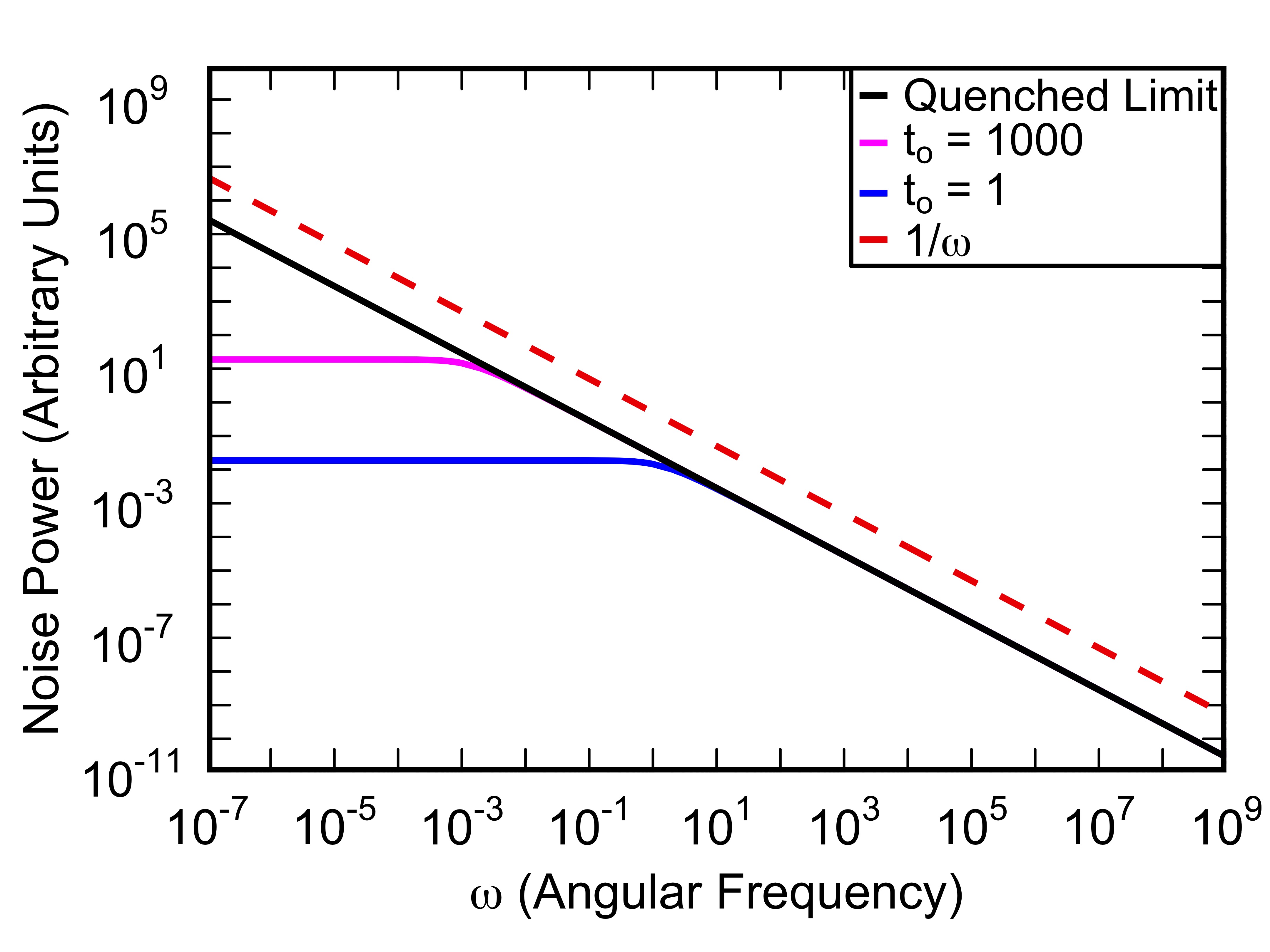}
\caption{Approximate noise spectra for a continuous distribution of rates, obtained from Eq.~(\ref{eq:powerspectrasimp}). The quenched limit ($t_o\rightarrow\infty$) is shown in black,  the magenta line corresponds to a rate change on average every $t_o=1000$ and the blue line corresponds to $t_o=1$. $\lambda=\log(\gamma_\text{max}/\gamma_\text{min})=\log(10^{12})$. The straight red dashed line  is proportional to $1/\omega$.   \label{fig:RandomFluctSpectraContLimit}}
\end{figure}

The overall behavior of the system is best seen with the limit $\gamma_\text{max}t_o\gg 1$ and $\gamma_\text{min}t_0\ll 1$. This eliminates the $1/\omega^2$ behavior at high frequencies as well as the low frequency flattening of the noise.  In that limit, $A(\omega)$ reduces to $-2\arctan(t_o\omega)$. $B(\omega)$ has a logarithmic dependence on $\omega$ and is small so it can be neglected entirely. This removes minor details of the frequency dependence of the noise. The expression for the power spectra is then greatly simplified:
\begin{eqnarray}\label{eq:powerspectrasimp}
\Psi(\omega)&\sim &\frac{1}{2}\frac{t_o\arctan(\omega t_o)/\lambda}{\omega t_o -\arctan(\omega t_o)/\lambda}\nonumber\\
&\sim & \frac{\arctan(\omega t_o)}{2\lambda\omega}
\end{eqnarray}
Now in the limit $\omega t_o\gg 1$, $\Psi(\omega)\rightarrow\pi/(4\lambda\omega)$. In this regime, the power spectra goes as $1/\omega$. In the limit $\omega t_o\ll 1$, the noise becomes independent of frequency, i.e., $\Psi(\omega)\rightarrow t_o/(2\lambda)$. The crossover between these two behaviors occurs around $\omega t_o\sim 1$ as expected. $\Psi(\omega)$, as approximated by Eq. (\ref{eq:powerspectrasimp}), is shown in Fig. \ref{fig:RandomFluctSpectraContLimit} for a few  values of $t_o$.

Thus we see that in order to observe 1/f noise in a given frequency range, the average duration $t_o$ of a fluctuation rate must exceed the inverse frequency 
of the lower limit of that range.

As a point of reference, we can put the rates used by Ahn {\it et al.}  \cite{Ahn2021} into our formulation Eq. (\ref{eq:RFTrelfuncchunked4}). Ahn {\it et al.} assumed thermally activated rates:
\begin{equation}\label{eq:AhnRate}
\gamma =\frac{1}{\tau} e^{-\frac{E}{k_bT_\text{sb}}}
\end{equation}
with a sub-bath temperature distribution $T_\text{sb}$ given by a Gaussian that was truncated to remove negative temperatures:
\begin{equation}\label{eq:AhnTdist}
f(T_\text{sb},T_{\rm avg},\sigma_\text{sb})=\xi\frac{e^{-(T_\text{sb}-T_{\rm avg})^2/2\sigma_\text{sb}^2}}{\sqrt{2\pi}\sigma_\text{sb}},
\end{equation}
where $T_{\rm avg}$ is the average sub-bath temperature,  $\sigma_\text{sb}^2$ the variance, and $\xi$ is a normalization factor that accounts for the truncation of the Gaussian. With the parameters used in this section, $\xi\sim 1$. 
Here we adopt Ahn's notation (except that we use $T_{\rm avg}$ rather than $T$). Eq.~(3) of Ahn {\it et al.}~\cite{Ahn2021}, written in terms of rates, is
\begin{equation}\label{eq:AhnNoise1}
\frac{S(\omega)}{\Delta^2\tau}=\int_0^{\infty}dT_\text{sb}f(T_\text{sb},T_{\rm avg},\sigma_\text{sb})\frac{2\tilde\gamma}{\omega^2\tau^2+\tilde\gamma^2},
\end{equation}
where $\tilde\gamma = \exp(-E/k_bT_\text{sb})$. $\Delta$ is the total variance of the signal produced by the switching events. Since our fluctuator jumps between $x=0$ and $x=1$, $\Delta^2=1/4$. Note that $\Delta$ is not the time interval that the system is at a given temperature; this use of $\Delta$ differs from that used earlier. Our Eq.~(\ref{eq:AhnNoise1}) differs from Ahn's Eq.~(3) by a factor of 2 which is likely due to Ahn's folding of the power spectra, i.e., Ahn assumes that the frequency is positive and includes the negative frequencies by multiplying the power spectrum by 2. We can explicitly include these quantities in Eq. (\ref{eq:AhnNoise1}):
\begin{equation}\label{eq:AhnNoise2}
\begin{split}
\frac{S(\omega)}{\Delta^2\tau} &= \int_0^{\infty}dT_\text{sb}\xi\frac{e^{-(T_\text{sb}/T_{\rm avg}-1)^2/2(\sigma_\text{sb}/T_{\rm avg})^2}}{\sqrt{\pi/2}\sigma_\text{sb}} \cdot \\
 & \quad \frac{2e^{-\frac{E}{k_bT_{\rm avg}}\frac{T_{\rm avg}}{T_\text{sb}}}}{\omega^2\tau^2+\left (e^{-\frac{E}{k_bT_{\rm avg}}\frac{T_{\rm avg}}{T_\text{sb}}}\right )^2}~.
 \end{split}
\end{equation}
A change of integration variable from $T_\text{sb}$ to $y=T_\text{sb}/T_{\rm avg}$ gives:
\begin{equation}\label{eq:AhnNoise3}
\frac{S(\omega)}{\Delta^2\tau}=4\int_0^{\infty}dy f(y,1,\tilde\sigma_\text{sb})\frac{2e^{-a/y}}{\omega^2\tau^2+\left(e^{-a/y}\right)^2}~,
\end{equation}
where $\tilde\sigma_\text{sb}=\sigma_\text{sb}/T_{\rm avg}$ and $a\equiv E/k_bT_{\rm avg}$. This expression from Ahn {\it et al.} \cite{Ahn2021} differs from our definition of the noise used in Eq.\ (\ref{eq:RFTrelfuncchunked4}) by a factor 2 due to a difference in normalization factors between Ahn {\it et al.} and us. Ahn {\it et al.}\ assumed that each temperature lasts for an infinite amount of time, i.e., $t_o=\infty$. We can generalize Eq.\ (\ref{eq:AhnNoise3}) to include a finite duration for temperature fluctuations by using Eq. (\ref{eq:RFTrelfuncchunked4}) which includes the factor $\Delta^2=1/4$:
\begin{equation}\label{eq:AhnNoise3Finite}
  \frac{S(\omega)}{\Delta^2\tau}
    =2t_o\cdot \frac{\mu_0^{'}(\omega)-(\mu_0^{'}(\omega))^2-(\mu_0^{''}(\omega))^2}{1-2\mu_0^{'}(\omega)+(\mu_0^{'}(\omega))^2+(\mu_0^{''}(\omega))^2}
\end{equation}
with:
\begin{equation}\label{eq:AhnRmu0Re}
\mu_0^{'}(\omega)=\int_0^{\infty}dy f(y,1,\tilde\sigma_\text{sb}) \frac{1+t_oe^{-a/y}}{(1+t_oe^{-a/y})^2+(\omega t_o)^2}
\end{equation}
\begin{equation}\label{eq:AhnRmu0Im}
\mu_0^{''}(\omega)= \omega t_o   \int_0^{\infty}dy f(y,1,\tilde\sigma_\text{sb}) \frac{1}{(1+t_oe^{-a/y})^2+(\omega t_o)^2}
\end{equation}
with {$t_o$ and
$1/\omega$ measured in units of $\tau$. Using the above equations with the values of Ahn {\it et al.}, namely, $a=E/k_bT_{\rm avg}=1$ and $\tilde\sigma_\text{sb}=0.3$, the effect of finite temperature fluctuations lifetimes is shown in Fig. \ref{fig:AhnGeneralized}. The noise for $t_o=10^{12}$ is indistinguishable from the quenched limit $t_o\rightarrow\infty$ shown in Fig. 1 of Ahn {\it et al.} \cite{Ahn2021} in the frequency range displayed. We see that shorter average durations of the temperature cause the curves to flatten off at higher frequencies where $\omega t_o\sim 1$. This reiterates our finding that low frequency 1/f noise requires very slow fluctuations. In the next section, we make estimates to see if this is physically reasonable in a 2DEG where Ahn {\it et al.} \cite{Ahn2021} assumed there would be temperature fluctuations in a sub-bath.

\section{Estimate of thermal relaxation time due to diffusion}
Experimentally measured charge noise exhibits 1/f behavior down to 1 Hz and even lower frequencies~\cite{Connors2019}, implying that if thermal fluctuations play an important role, the relaxation time of these thermal fluctuations must be significantly longer than the relaxation time of the fluctuators coupled to the quantum dots, i.e., the thermal fluctuations must last at least a few seconds or longer. To see whether this is reasonable, we can estimate the duration of temperature fluctuations in the 2DEG. Temperature fluctuations in a sub-bath imply spatial inhomogeneities in the temperature of 2DEG. Thermal diffusion would smooth out these inhomogeneities. 
\begin{figure}[h]
\centering
\includegraphics[width=8.5cm]{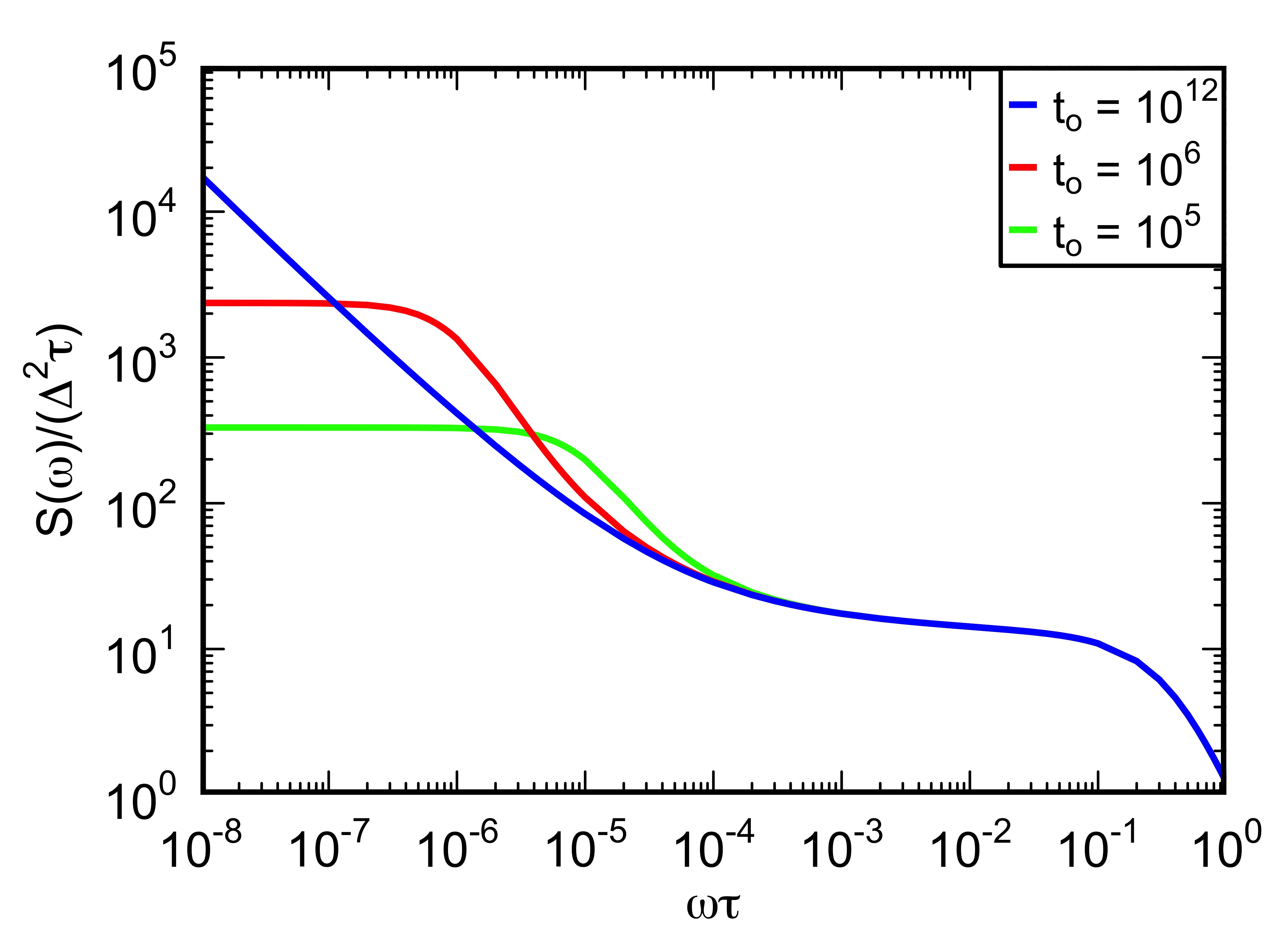}
\caption{Noise power spectra vs. $\omega\tau$ on a log-log plot for three different average switching times: $t_o=10^{12}$ (blue), $t_o=10^6$ (red), and $t_o=10^5$ (green). These plots show the effect of finite temperature fluctuations lifetimes on the noise spectra of Ahn {\it et al.}.   \label{fig:AhnGeneralized}}
\end{figure}
Long-lived fluctuations are difficult to achieve in typical electronic systems without some kind of activated behavior. An exception exists when a quantity satisfies a conservation law. Then long relaxation times can exist for large scale fluctuations. This possibility was investigated by Voss and Clark \cite{voss76_flick_noise} for energy fluctuations (equivalently temperature fluctuations via $\Delta E=C_v\Delta T$ where $C_v$ is the heat capacity in metal films). Since the slowest relaxation occurs for the largest spatial fluctuations, the dimensions of the system introduce key frequencies in the problem: $f_i=D/(\pi l_i^2)$ where $D$ is the thermal diffusion constant and $l_i$, with $i=x,y,z$, are the dimensions of the system. There are no fluctuations that will last longer than $1/f_i$ and therefore all that is needed is to evaluate $f_i$ for the largest dimension which is about 1 $\mu$m for the 2DEG in the work of Connors et al. \cite{Connors2019}. (In~\cite{voss76_flick_noise}, the dimensions of the conductor were quite large, of the order of millimeters,  and together with a  diffusion constant equal to  $D\sim 2\cdot 10^{-5}$ m$^2$/s, produced minimal frequencies of the order of 1 Hz or less.) To estimate the lowest frequency in the present system, the diffusion constant $D$ must be evaluated at low temperatures for the 2DEG. In what follows, the diffusion constant is expressed in terms of quantities measured in the Si/SiGe 2DEG of ref. \cite{Connors2019}. 

In a diffusive regime (which is assumed to be the case here), the diffusion constant $D$ is related to the thermal conductivity $\kappa$ via
\begin{equation}\label{eq:diff2D}
  D=\frac{\kappa}{C},
\end{equation}
where $\kappa$ is the thermal conductivity in W/K (in two dimensions) and $C$ is the specific heat in $\text{J}/(\text{K}\cdot \text{m}^2)$. For the 2DEG, the specific heat is given by \cite{Ahn2021}
\begin{equation}\label{eq:C2D}
  C=\frac{\pi m k_b^2T}{3\hbar^2},
\end{equation}
where the carrier's mass is $m=0.19m_e$, and $m_e$ is the mass of the electron.

The thermal conductivity is rarely available experimentally. Using the Wiedemann–Franz law, it is possible to relate the thermal conductivity to the electric conductivity $\sigma$:
\begin{equation}\label{eq:WFLaw}
  \frac{\kappa}{\sigma T} = \frac{\pi^2}{3}\left(\frac{k_b}{e}\right)^2,
\end{equation}
where $e$ is the electric charge. The final piece is to compute the electrical conductivity which is given by
\begin{equation}\label{eq:conductivity2D}
  \sigma = e\cdot n\cdot \mu,
\end{equation}
where $n$ is the carrier number density and $\mu$ is the carrier mobility which is usually available. Combining Eqs.~(\ref{eq:diff2D}), (\ref{eq:C2D}), (\ref{eq:WFLaw}) and (\ref{eq:conductivity2D}) gives the diffusion constant in terms of measured quantities:
\begin{equation}\label{eq:diffusion2D}
  D=\frac{\pi n \mu \hbar^2}{e m}.
\end{equation}
Using the values $n=2.2\cdot 10^{15}$ 1/m$^2$ and $\mu=16$ m$^2$/(Volt s) reported in ref. \cite{zajac15_recon_gate_archit_si_quant_dots}, Eq. (\ref{eq:diffusion2D}) gives $D=1.4\cdot 10^{-2}$ m$^2$/s.

Using this value for $D$, we estimate the lowest frequency in the problem to be
\begin{equation}
  f_{\rm min}=\frac{D}{\pi l^2}=\frac{1.4\cdot 10^{-2}}{\pi (10^{-6})^2}  \ \textrm{Hz} = 4.5 \ \textrm{GHz}.
\end{equation}
It is therefore highly unlikely that a 2DEG bath can satisfy the assumption underlying the calculation of Ahn {\it et al}. \cite{Ahn2021}.

We can obtain a slightly different estimate in a different 2DEG (AlN/GaN) system where the Wiedemann-Franz law was verified \cite{abou-hamdan20_tunin_elect_therm_conduc_two}. This paper measured the specific heat as well as thermal and electrical conductivity of the 2DEG. The thermal conductivity has the form $\kappa=90 T/275$ W/(K m) (where $T$ is temperature) while the specific heat is $C=0.05 T$ in J/(kg K). (Note that the units are appropriate for 3D quantities; the paper \cite{abou-hamdan20_tunin_elect_therm_conduc_two} measured the thickness of the electron gas for the conversion.) The diffusion constant is then $D=\kappa/\rho C$ where $\rho$ is the density of the material (GaN) and is equal to 6150 kg/m$^3$. The temperature dependences cancel out and the diffusion constant is found to be approximately $10^{-3}$ m$^2$/s. This value is slightly smaller than what has been estimated for the Si/SiGe case but does not change the conclusion. 

Finally, the diffusion constants obtained above are significantly larger than those used by Voss and Clarke in the context of metal thin films ($D\sim 2\cdot 10^{-5}$ m$^2$/s). However, even with such a diffusion constant, the conclusion remains unchanged (the lowest frequency drops to 6 MHz). 

\section{Conclusions}
We have considered a quantum dot whose charge noise is determined by a fluctuator coupled to a thermal bath with a fluctuating temperature. We have used Monte Carlo simulations of an Ising spin in a fluctuating temperature bath as well as analytic calculations of a two-state fluctuator with random switching rates to determine the noise spectrum  of this  two-level fluctuator.
We find that a 1/f noise spectrum at a given frequency $f_0$ requires that the frequencies of the thermal fluctuations must be comparable to $f_0$. However, our estimate of the lowest temperature fluctuation frequency is a few GHz in a 2DEG which is inconsistent with 1/f noise observed at frequencies below 1 Hz. In short, 
to obtain 1/f noise that extends over several decades in frequency from a model based on temperature fluctuations requires fluctuations with an unphysically long duration.

\acknowledgments
This work was supported in part by the National Science Foundation (NSF) through the University of Wisconsin Materials Research Science and Engineering Center (Grant No.~DMR-1720415) and by the Australian Research Council (Project No.~DP210101608).


%
\end{document}